\documentclass[%
  reprint,
superscriptaddress,
showpacs,preprintnumbers,
 amsmath,amssymb,
aps,
prl,
floatfix,
]{revtex4-1}

\usepackage[utf8]{inputenc}

\usepackage{graphicx}
\usepackage{tikz}
\usepackage{pgfplots}
\pgfplotsset{compat=newest}
\usetikzlibrary[pgfplots.groupplots]
\usetikzlibrary{arrows}
\usetikzlibrary{decorations}
\usepgfplotslibrary{external}
\usetikzlibrary{backgrounds}
\usetikzlibrary{chains}
\usetikzlibrary{calc}
\tikzstyle{simplefleche}=[->,>=latex,thin]
\tikzstyle{doublefleche}=[<->,>=latex,thin]
\tikzstyle{dotdoublefleche}=[<->,>=latex,thin,densely dotted]
\tikzstyle{doubleflechebar}=[<->,>=|,thin]

\usepackage{bm}

\usepackage{empheq}
\usepackage{xspace}
\usepackage{color}
\usepackage{amssymb}
\usepackage{hyperref}
\hypersetup{
bookmarksnumbered = true,
unicode=false,          
pdftoolbar=true,        
pdfmenubar=true,        
pdffitwindow=false,     
pdfstartview={Fit},    
pdftitle={Power_Stroke},    
pdfauthor={Author},     
pdfsubject={Subject},   
pdfcreator={Creator},   
pdfproducer={Producer}, 
pdfkeywords={keyword1} {key2} {key3}, 
pdfnewwindow=true,      
colorlinks=black,       
linkcolor=black,          
citecolor=black,        
filecolor=black,      
urlcolor=black,           
pdfdisplaydoctitle = true
}

\newcommand{\mean}[1]{\langle #1 \rangle}
\usepackage[normalem]{ulem}
\usepackage{color}
\definecolor{myorange}{rgb}{0.9568,0.4941,0.1961}
\definecolor{myred}{rgb}{0.9098,0.1294,0.2078}
\definecolor{myblue}{rgb}{0.0352,0.4981,0.6509}
\definecolor{mygreen}{rgb}{0.2235,0.6353,0.2588}

\usepackage{marvosym}

\begin{document}

\preprint{APS/123-QED}

\title{Muscle as a Metamaterial Operating Near a Critical Point}

\author{M. Caruel}
\affiliation{Inria, 1 rue Honor\'{e} d'Estienne d'Orves,
91120 Palaiseau, France
}
\affiliation{
LMS,  CNRS-UMR  7649,
Ecole Polytechnique, 91128 Palaiseau Cedex,  France
}
\author{J.-M. Allain}
\affiliation{
LMS,  CNRS-UMR  7649,
Ecole Polytechnique, 91128 Palaiseau Cedex,  France
}
\author{ L. Truskinovsky}
\email{trusk@lms.polytechnique.fr}
\affiliation{
LMS,  CNRS-UMR  7649,
Ecole Polytechnique, 91128 Palaiseau Cedex,  France
}
\date{\today}

\begin{abstract}

The passive mechanical response of skeletal muscles at fast time scales is dominated by long range interactions inducing cooperative behavior without breaking the detailed balance. This leads to such unusual ``material properties'' as negative equilibrium stiffness and different behavior in force and displacement controlled loading conditions. Our fitting of experimental data suggests that ``muscle material'' is finely tuned to perform close to a critical point which  explains large fluctuations observed in muscles close to the stall force.

\end{abstract}

\pacs{Valid PACS appear here}
\maketitle

\begin{figure}[htbp]
	\centering
	\includegraphics{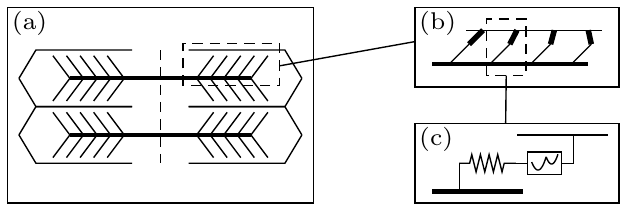}
	\caption{Schematic structure of the three layers of organization inside a sarcomere: (a) global architecture with domineering parallel links; (b)  structure of an elementary contractile unit shown in more detail in Fig.~\ref{fig:cat_model}; (c) individual attached cross-bridge  represented by a bistable element in series with a shear spring. }
	\label{fig:sarcomere_topology}
\end{figure}

Active behavior of skeletal muscles is associated with  time scales  of about 30 ms \cite{Alberts:2007vj,*Howard_2001}.  At shorter times ($\sim$1 ms) muscles  exhibit a nontrivial \emph{passive} response: if  a tetanized muscle is suddenly extended, it comes loose, and if it is shortened, it tightens up with apparently no involvement of Adenosine Triphosphate (ATP) \cite{Huxley_1971,*Ford_1977,*Lombardi_1992,*Piazzesi_1992,*Linari_2009}. This unusual  mechanical behavior,  associated with the unfolding of the attached myosin cross-linkers   (cross-bridges), qualifies  muscles as metamaterials \cite{Nicolaou:2012cf}. As we argue below, an important factor in this behavior is the dominance of parallel connections with multiple shared links  entailing cooperative effects; see Fig.~\ref{fig:sarcomere_topology}. Similar mean-field  coupling can be found in many hierarchical biological systems \cite{Kometani_1975,*Desai_1978,*Dauxois:2003um}; in particular, it plays a crucial role in cell adhesion, where individual binding elements interact through a common elastic background \cite{Erdmann_2007}.

Interaction-induced synchronization during muscle contractions  reveals itself through macroscopic fluctuations and spatial inhomogeneities
\cite{Placais_2009,*Pavlov:2009jv,*Shimamoto_2009,*Holzbaur:2010fh,*Ishiwata_2011}.
 In ratchet-based  and chemomechanical models such collective behavior is usually attributed to breaking of the detailed balance \cite{Julicher_1995,*Julicher_1997a,*Guerin_2010,*Guerin_2011}, with
 long range interactions  entering the problem implicitly as a force dependence of the chemical rates
\cite{Duke_1999,*Duke_2000,*Smith_2008,*Erdmann:2012cr}. However, the cooperative behavior of myosin cross-bridges can be detected during short time force recovery \cite{Armstrong:1966ta,*Edman_1988,*Granzier_1990,*Edman_2001}, and therefore  the origin of synchronization should be within reach of models disregarding  disequilibrium and activity. In this Letter, we  show that  already  equilibrium response of ``muscle material'' is  associated with highly synchronized behavior at the microscale which explains its unusual passive response.

In particular, we show that an order-disorder  phase transition is displayed by the celebrated Huxley-Simmons (HS) model
 \cite{Huxley_1971} if, instead of physiological isometric loading conditions (length clamp) also known as a \emph{hard device}, one considers  isotonic (load clamp)  loading conditions or a \emph{soft device}. While a considerable difference in  behavior of a muscle loaded in these two different ways can be deduced from experimental data \cite{Piazzesi_2002,*Reconditi_2004,*Piazzesi_2007},   the origin of the disparity has been so far unexplained. We argue that behind it is a nonequivalence of equilibrium  ensembles ubiquitous  in systems with long range interactions \cite{Campa:2009wz}.

Most remarkably, we find that a careful parameter fitting places the actual skeletal muscle almost exactly into a ferromagnetic Curie  point. This agrees with the observation \cite{Huxley_1971} that the effective stiffness of skeletal muscles associated with fast force recovery is close to zero in the state of isometric contractions and strongly suggests that muscles are finely tuned to perform near marginal stability. Other experimentally observed manifestations of criticality include kinetic slowing down and large scale macroscopic fluctuations near the stall force conditions \cite{Armstrong:1966ta,*Edman_1988,*Granzier_1990,*Edman_2001}.

We demonstrate the robustness of our predictions by comparing the HS  model, which we interpret as a hard spin description, with a regularized (RHS) model where filament elasticity is  taken into consideration and conventional spins are replaced by elastic  snap springs.

\begin{figure}
 \centering
\includegraphics{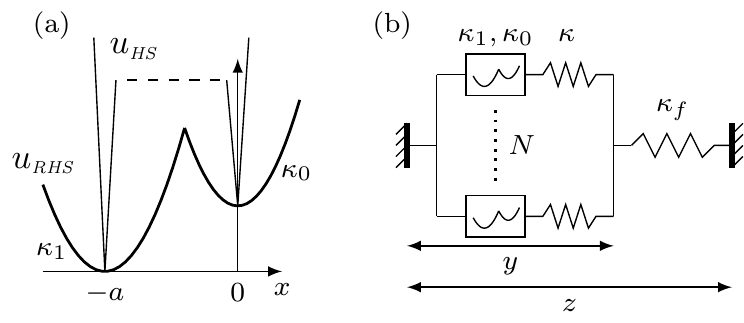}
 \caption{Elementary contractile unit. (a)  Energy of the bistable (power-stroke) element: HS model (thin line) and RHS model (thick line); (b)  \( N \) cross-bridges in hard device. In the HS model, \( \kappa_{0},\kappa_{1},\kappa_{f}=\infty \) and  \( y=z \).}
\label{fig:cat_model}
\end{figure}

\emph{The HS model.}--- We consider a prototypical model of a half-sarcomere  with  $N$ attached cross-bridges arranged in parallel \cite{Huxley_1971}; at time scales of fast force recovery  $N$ can be considered constant \cite{Lombardi_1992,Reconditi_2004}. Each cross-bridge is modeled as a series connection of a bistable spin unit and a linear (shear) spring; see Fig.~\ref{fig:cat_model}. We use  dimensionless variables with the power-stroke size \( a \) as a unit of displacement and  \( \kappa a^{2} \) as the unit of energy, where \(\kappa\) is the stiffness of the series spring. Then the spin variable takes values $x_{i}=0$ (pre-power-stroke state) and $x_{i}=-1$  (post-power-stroke state), and  the total energy per particle in the hard device  is $v(\boldsymbol{x},z) = (1/N) \sum [(1+x_{i})v_0 + (1/2)\left( z-x_{i} \right)^{2}]$,
where $v_0$ is the energetic bias of the pre-power-stroke state. In this formulation, the HS model describes the simplest paramagnetic spin system. 

 At finite temperature \( \theta \), the equilibrium behavior of this system is characterized by the free energy per particle
\(
	\hat{f}(z,\beta)=-[1/ (N \beta)]\ln \int
	\exp\left[-\beta Nv(\boldsymbol{x},z)\right]d\boldsymbol{x}
\)
where $\beta = \kappa a^{2}/(k_{b}\theta)$. At fixed   $p=-(1/N) \sum x_{i}$, representing the fraction of cross-bridges in the post-power-stroke state, the macroscopic ( $N\to\infty$ ) free energy   takes the form
\begin{equation*}
	f =  p\biggl[\frac{1}{2}( z+1)^{2}\biggr] + (1-p)\biggl[ \frac{1}{2}z^{2} + v_{0}\biggr]  + \frac{1}{\beta}S(p)
\end{equation*}
where  $S(p)=  p\log(p) + (1-p)\log(1-p)$.
The function $f(p)$ is  always convex with a minimum at
$\hat{p}(z,\beta) =  1/2 -(1/2)\tanh[(\beta/2)(z-v_{0}+1/2)]$. The equilibrium tension per cross-bridge is then $t = (z+\hat{p} )$, which is exactly the formula found by Huxley and Simmons. The equilibrium free energy is
	\begin{equation*}
		\hat{f} = - \frac{1}{\beta}\ln\!\biggl\{\exp\biggl[-\frac{\beta}{2} (z+1)^2\biggr]\,+\,\exp\biggl[-\beta\biggl(\frac{z ^{2}}{2} +v_0\biggr)\biggr]\biggr\}
	\end{equation*}
 and the susceptibility (stiffness) is $ t'(z)=\hat{f}''(z)$.  The function \( \hat{f}(z) \) is convex for  \( \beta<4 \) and is nonconvex for \( \beta>4 \)  exhibiting a range with negative equilibrium stiffness (metamaterial behavior).   If we take the values from \cite{Huxley_1971}, \( a=8 \) nm and  \( \kappa a^{2}/2 = 2 k_{b}\theta \),  we obtain  $\beta = 4$, which corresponds to zero stiffness at the state of isometric contractions $z_0=v_{0}-1/2$.

In the soft device setting, not studied by Huxley and Simmons, the  energy becomes  \( w(\boldsymbol{x},z,t) = v(\boldsymbol{x},z) - tz \), where $t$ is the applied force per particle.
Now the variable $z$ plays the role of an internal parameter whose adiabatic elimination produces a Curie-Weiss mean-field potential depending on $ \left(\sum x_{i}\right)^2$. The equilibrium Gibbs free energy is now
\(
	\hat{g}(t,\beta)=-[1/ (N \beta)]\ln \int
	\exp\left[-\beta Nw(\boldsymbol{x},z,t)\right]d\boldsymbol{x}dz.
\)
At fixed $p$  we obtain in the thermodynamic limit
\begin{equation*}
	g =  -\frac{1}{2} t^{2} + pt +(1-p) v_{0}+ \frac{1}{2}p(1-p) + \frac{1}{\beta}S(p),
\end{equation*}
where the ``regular solution'' term is responsible for cooperative (ferromagnetic) behavior.

\begin{figure}[htbp]
	\centering
	\footnotesize
		\includegraphics{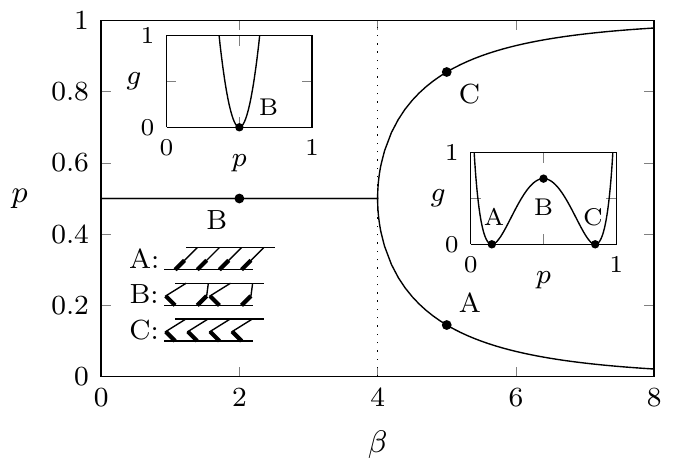}
		\caption{Bifurcation diagram for the HS model placed in a soft device showing synchronized states (A and C) and disordered state (B).  Solid lines shows minima of the Gibbs free energy at \( t = 1 \). The critical point is located at \( \beta=4 \). }
\label{fig:hs_bifurcation}
\end{figure}

In Fig.~\ref{fig:hs_bifurcation}, we show the position of the minima of $g(p)$  when $t$ is chosen to ensure that in the paramagnetic phase \(\hat{p}(t,\beta)=1/2\). In the disordered (high temperature) state all cross-bridges are in random conformations, while in  the  ordered (low temperature) state the system exhibits coherent fluctuations between   post-power-stroke and pre-power-stroke configurations. These fluctuations describe \emph{temporal microstructures} responsible for the plateau in the force-elongation relation  \(\hat{z} = t - \hat{p}\),  where  \( \hat{p} \) is a solution of \( t-\hat{p}+1/2-v_{0}+ (1/\beta)\ln[\hat{p}/(1-\hat{p})]=0 \).

The equilibrium Gibbs energy is  concave because \( \hat{g}''(t) = -1-\beta N \langle\left(p-\hat{p}\right)^{2}\rangle\leq 0 \), so in the soft device the  stiffness is always positive. Since in the hard device the  stiffness is sign indefinite, the two ensembles are not equivalent.
This is expected for systems with strong long range interactions that are inherently nonadditive \cite{Campa:2009wz,Campa:2007hx,*Ellis:2000df}. Negative stiffness in the hard device HS model has been known for a long time \cite{Huxley_1971,Hill_1975a,*Hill_1975b,Duke_2000,Marcucci_2010,*Marcucci_2010a}; however, it was not previously associated with the particular internal architecture of muscle material.

As we have already mentioned, the original HS fit of  experimental data \cite{Huxley_1971} places the system exactly into the \emph{critical state} (Curie point). In this state the correlation length diverges and  fluctuations become macroscopic, which is consistent with observations  at stall force conditions \cite{Telley_2006b,*Pavlov:2009jv,*Pavlov:2009uw,*Serizawa:2011kj}.
This suggests that skeletal muscles, as many other biological systems,  may be tuned to criticality. The  proximity to the critical point would then be the result of either evolutionary or functional self-organization.  The  marginal stability of the critical state  allows the system   to amplify  interactions, ensure strong feedback, and achieve considerable robustness in front of random perturbations. In particular, it is  a way  to quickly switch back and forth between highly efficient synchronized stroke and stiff behavior in the desynchronized state.

\begin{figure}[htbp]
	\centering
	\includegraphics{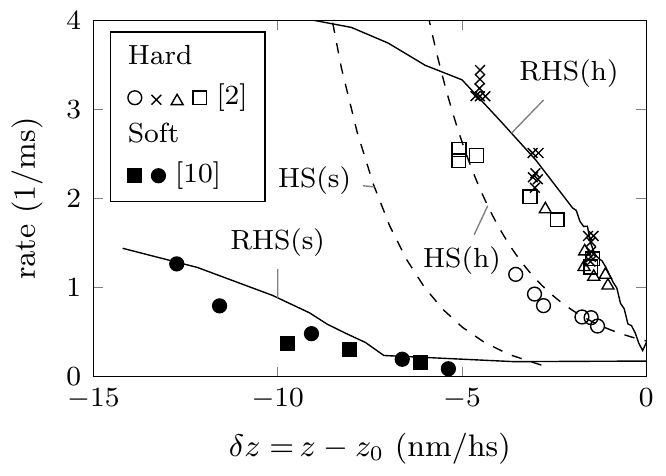}
	\caption{Recovery rates in hard and soft devices. Symbols: Postprocessing of experimental data; see \cite{Piazzesi_2002,*Reconditi_2004,*Piazzesi_2007}. Open symbols, hard device; filled symbols, soft device. Dashed lines: HS model in hard (h) and soft (s) devices; parameters are taken from \cite{Huxley_1971}. Solid lines: RHS model in hard (h) and soft (s) devices obtained from stochastic simulations;  parameters have been fit to experimental data: \( \lambda_{1} = 0.41 \), \( \lambda_{0} = 1.21 \), \( \lambda_{f}=0.72 \), \( l=-0.08 \), \( N=112 \), \( \beta=52 \) (\( \kappa =2 \) pN/nm, \( a=10 \) nm, \( \theta = 277.13 \) K), \( z_{0}=4.2 \) nm/hs.}
	\label{fig:cat_fitted_rates}
\end{figure}

The ensembles nonequivalence in the HS model has also a kinetic signature. Experiments  on quick recovery reveal  that   muscle fibers  react to load steps much slower than to length steps \cite{Reconditi_2004,*Piazzesi_2007,*Linari_2009}.
 This agrees with our model, where coherent response (in isotonic conditions) is always slower than disordered response (in isometric conditions). 

Indeed, by using Kramers approximation Huxley and Simmons  obtained in a hard device 
the kinetic equation \(\dot{\mean{p}} = -k_{-}\mean{p} + k_{+}(1-\mean{p})\), where \( \mean{p} \) is the average over ensemble. The constants \( k_{+}, k_{-} \) satisfy the detailed balance  \(k_{+}/k_{-}=\exp\left[-\beta(z-v_{0}+1/2)\right] \), and
 the recovery rate is  $1/\tau= k_{-}\{1+\exp[-\beta(z-v_{0}+1/2)]\}$. 
In a soft device we may use the same model with \( z = t-\mean{p} \), which accounts for force-dependent chemistry and introduces nonlinear feedback. The characteristic rate around a given state $\mean{p}$ is then \( 1/\tau  = k_{-}\{1+[1-\beta(1-\mean{p})]\exp[-\beta(t-\mean{p}-v_{0}+1/2)]\} \). When $\mean{p}$ is small, $t-\mean{p}> z$, and the relaxation in a soft device is  slower than in a hard device.  In Fig.~\ref{fig:cat_fitted_rates}, we show the rates obtained from the HS model; in the case of a soft device, the nonlinear kinetic equation was solved numerically for the duration 10 ms. We see that in a soft device the rates are indeed slower than in a hard device, however, the experimental measurements are not matched quantitatively.

\emph{The RHS model.}--- To test the robustness of the HS mechanism of synchronization and to achieve  quantitative agreement with kinetic data, we now consider a natural regularization of the HS model. First, following \cite{Marcucci_2010} we replace hard spins by soft spins described by a piecewise quadratic  double well potential --~see Fig.~\ref{fig:cat_model}(a)--~
\begin{equation*}
 u_{\mbox{\tiny\itshape  RHS}}(x) =
\begin{cases}
 \frac{1}{2}\lambda_0(x)^{2} +  v_{0}  & \text{if $ x> l $,}\\
  \frac{1}{2}\lambda_1(x + 1)^{2}  & \text{if $ x\leq  l $,}
\end{cases}
\end{equation*}
where \( \lambda_{1} = \kappa_{1}/\kappa \), \( \lambda_{0} = \kappa_{0}/\kappa \). Second, we introduce a mixed device (mimicking myofilament elasticity
\cite{Huxley_1994,*Wakabayashi_1994,Julicher_1995}) by adding to our parallel bundle a series spring. The resulting energy  per cross-bridge in a hard device is
\begin{equation*}
 v(\boldsymbol{x},\!y;\!z)=\frac{1}{N} \sum_{i=1}^{N} \big[u_{\mbox{\tiny\itshape  RHS}}(x_{i}) + \frac{1}{2}( y-x_{i})^{2}\big]+\frac{\lambda_{f}}{2}(z-y)^{2},
 \label{eq:hs_energy_nodim_hd 1}
\end{equation*}
where $y$ is a new internal variable and \( \lambda_{f} = \kappa_{f}/(N\kappa) \); see Fig.~\ref{fig:cat_model}(b). It is clear that our lump description  of filament elasticity  misrepresents  short range interactions  \cite{Ford:1981wd,*Mijailovich_1996,*DeGennes_2001}; however, this should not affect our main results \cite{Nagle:1970kr,*Kardar:1983tq}.

To study the soft device case we must again consider the energy $w(\boldsymbol{x},y,z,t)=v(\boldsymbol{x},y,z)-tz$, where $t$ is the applied force per cross-bridge. A transition from hard to soft   ensemble  is made by  taking  the limit  $ \lambda_{f}\rightarrow 0$, $z \rightarrow \infty$  with  $ \lambda_{f}z \rightarrow t$.  At finite $\lambda_{f}$  the RHS model can be viewed as a version of the  mean-field $\varphi^4$ model studied  in   \cite{Desai_1978,*Dauxois:2003um,Campa:2009wz}. 

The HS model is a limiting case of the RHS model with  $\lambda_{1,0}\rightarrow\infty$ and $\lambda_{f}\rightarrow\infty$. The first of these limits  allows one to replace continuous dynamics by jumps and use the language of chemical kinetics; however, it also erases information about the barriers; see \cite{Marcucci_2010}. The second limit eliminates the Curie-Weiss (mean-field) interaction among individual cross-bridges  at fixed $z$, and that is why the synchronized behavior was overlooked in \cite{Huxley_1971}. 

\begin{figure}[htbp]
	 \centering
	\footnotesize
	 \includegraphics{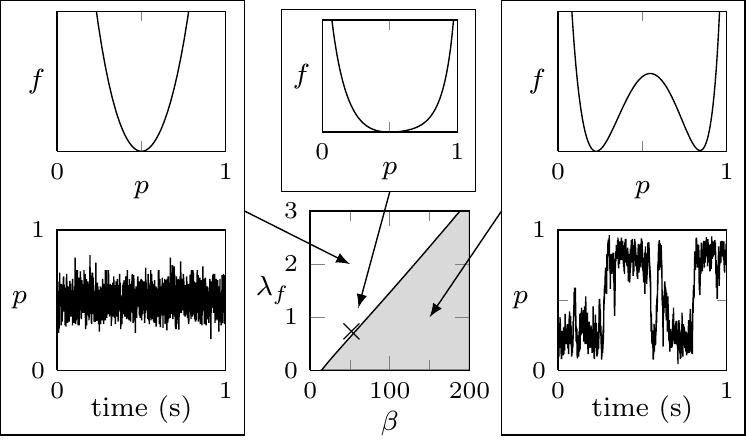}
	\caption{
	Phase diagram for the RHS model in a hard device with \( z \) selected to ensure that \( \langle p\rangle=1/2 \) at each point $(\beta,\lambda_{f})$. In the shaded region, the function \( f (p) \) is nonconvex which leads to coherent fluctuations. Outside this region, fluctuations are not synchronized. The cross indicates an almost critical configuration with realistic parameters (used in Fig. \ref{fig:cat_fitted_rates}).}
	\label{fig:cat_phase_diagram_hd}
\end{figure}
Equilibrium behavior in the RHS model can be again described analytically, because it is just a redressed HS model.  In the limit \( \lambda_{f}\to\infty \)  the function \( f (p) \) is convex as in HS model, while at finite values of \( \lambda_{f} \) it is now nonconvex.
This shows that in the RHS model the account of filament elasticity brings about phase transition (and bistability) also in the hard device.  

The bistable nature of the macroscopic free energy  in both soft and hard devices implies that the system can be in two coherent states, and  therefore within a large set of half-sarcomeres one should expect observable spatial inhomogeneities. This prediction is in agreement with ubiquitous ``off-center'' displacements of \emph{M} lines recorded during isometric contractions \cite{Pavlov:2009jv}.

\begin{figure}[htbp]
	\centering
	\includegraphics{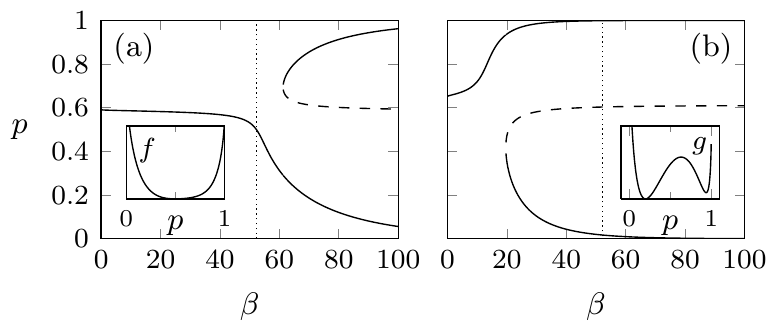}
	\caption{Bifurcation diagram for nonsymmetric  RHS model with realistic parameters: (a)  hard device with $z=0.37$; (b)    soft device with $t=0.21$; this loading  secures that $\langle p\rangle=1/2$ for \( \beta=52 \). Parameters are as in Fig. \ref{fig:cat_fitted_rates}.  Inset (a) corresponds to \( \beta = 52 \),   inset (b) to \( \beta=25 \).
	}
	\label{fig:cat_bifurcation_diagrams}
\end{figure}

The phase diagram showing the role of filament elasticity in hard device is shown in Fig.~\ref{fig:cat_phase_diagram_hd}. The  dependence of the critical temperature on \( \lambda_{f}\) suggests that actomyosin systems can control the degree of cooperativity by tuning the internal stiffness;  likewise, variable stiffness of the loading device may be used in experiments to either activate or deactivate  the collective behavior. Notice that the realistic choice of parameters again selects a near critical state; the exact criticality is compromised  since the symmetry between the pre- and post-power-stroke states  is now broken  (as $\lambda_{1} \neq \lambda_{0}$) and the phase transition becomes weakly first order; see Fig.\ref{fig:cat_bifurcation_diagrams}.

A behavior  similar to  our synchronization has been also observed in the models of passive adhesive clusters,  where the elastic feedback   appears as strain- or force-dependent chemistry \cite{Erdmann_2007}.  Given that the  two  systems exhibit almost identical cooperative behavior, we expect criticality to be also a factor in the operation of focal adhesions.

\begin{figure}[htbp]
	\centering
	\includegraphics{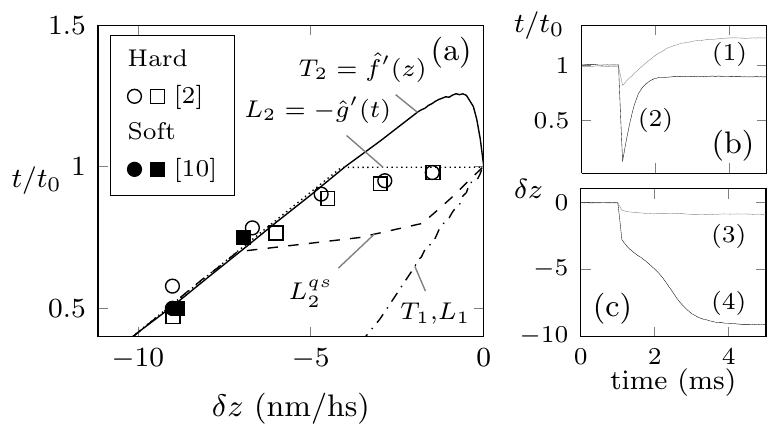}
	\caption{(a) States attainable during quick recovery: solid line, $T_2$; dotted line, $L_2$; dashed line, $L^{qs}_2$ corresponds to quasistationary states; dash-dotted line, $T_1$  and $L_1$. Symbols show experimental points for hard (open) and soft (filled) devices;  see \cite{Piazzesi_2002,*Reconditi_2004,*Piazzesi_2007}. In a hard device, the equilibrium \( T_{2} \) curve coincides with the results of stochastic simulations.  In a soft device, the equilibrium \( L_{2} \) curve differs from  the simulation results at 10 ms (\( L_{2}^{qs} \) curve). Averaged trajectories after abrupt loading at  $1$ ms: (b)  in a hard device and (c)  in a soft device. Curve (1): \( \delta z = -1 \) nm/hs; curve (2): \( \delta z = -5 \) nm/hs; curve (3): \( t/t_0=0.9 \); curve (4): \( t/t_0=0.5 \) with \( t_{0} = 0.21 \). Other parameters are as in Fig.\ref{fig:cat_fitted_rates}.}
	\label{fig:cat_t1_t2}
\end{figure}

The two ensembles, soft and hard, remain inequivalent in the RHS model. Thus, in the soft device   the equilibrium Gibbs free energy \( \hat{g} \) is  concave since 
 \( \hat{g}''(t) = -1/\lambda_{f} - \beta N \langle(y-\langle y\rangle)^{2}\rangle\leq 0 \)  which means that the stiffness is always positive.  Instead in the hard device  
\( \hat{f}''(z) = \lambda_{f}[1-\beta N \langle(y- \langle y\rangle)^{2}\rangle] \),
and the stiffness can be both positive  and negative. While negative stiffness should be a characteristic feature of realistic half-sarcomeres (see Fig.~\ref{fig:cat_t1_t2}), it has not been observed  in experiments on whole myofibrils. The reason may be that in myofibrils a single half-sarcomere is never loaded in a hard device. The effective dimensionless temperature  may also be higher because of the quenched disorder, and the stiffness may be smaller due to  nonlinear elasticity. One can also expect the unstable half-sarcomeres to be stabilized actively through processes involving ATP hydrolysis \cite{Vilfan_2005}.

To study kinetics in the RHS model we perform  direct numerical simulations by using a Langevin thermostat. We assume that the macroscopic variables $y$ and $z$ are fast and are always mechanically equilibrated which is not an essential assumption. The response of the remaining variables $x_i$ is governed by the system
$ dx_{i} = b(x_{i})dt + \sqrt{2\beta^{-1}}dB_{i}$,
where the drift is $ b (\boldsymbol{x},z) = -u'_{\mbox{\tiny\itshape  RHS}} (x_{i}) + (1+\lambda_{f})^{-1}( \lambda_{f}z +  N^{-1}\sum x_{i}) - x_{i} $ in a hard device and $ b (\boldsymbol{x},t) = -u'_{\mbox{\tiny\itshape  RHS}} (x_{i}) + t + N^{-1}\sum x_{i} - x_i $ in a soft device. In both cases the diffusion term $dB_i$ represents a standard Wiener process. 

 In Fig.~\ref{fig:cat_t1_t2}, we show the results of stochastic simulations imitating quick recovery experiments \cite{Huxley_1971}. The system, initially in thermal equilibrium at fixed \( z_{0} \) (or \( t_0 \)), was perturbed by applying fast (\( \sim100\ \mu \)s) length (or load) steps with various amplitudes.   In a soft device the system was not able to reach  equilibrium  within the experimental time scale. Instead, it  remained trapped in a quasistationary (glassy) state because of the high energy barrier associated with collective power stroke. Such kinetic trapping  which fits the pattern of two-stage dynamics exhibited by systems with strong long range interactions   \cite{Bouchet:2010gf,*Nardini:2012vs,Campa:2009wz} may explain the failure to reach equilibrium in experiments reported in \cite{Edman_1988,*Granzier_1990,*Edman_2001}.
 In the hard device case, the cooperation among the cross-bridges is weaker and kinetics is much faster,  allowing the system to reach equilibrium at the experimental time scale. A quantitative comparison of the rates obtained in our simulations with experimental values  (see   Fig.~\ref{fig:cat_fitted_rates})  shows  that the RHS model reproduces the kinetic data in both hard and soft ensembles rather well. 

In conclusion,  we mention that the prototypical nature of our model implies that \emph{passive collective behavior} should be a  property common to general cross-linked actomyosin networks. We have shown that the degree of cooperativity in such networks can be strongly affected by elastic stiffness of the filaments.  This suggests that a generic system of this type can be tuned to criticality by an actively generated  prestress \cite{Sheinman:2012gi}.

 \begin{acknowledgments}
	We thank D. Chapelle, J.-F. Joanny, K.Kruse, and V. Lombardi for insightful comments. M. Caruel thanks Monge PhD fellowship from Ecole Polytechnique for financial support.
\end{acknowledgments}

\bibliography{biblio_short}
\bibliographystyle{apsrev4-1}
\end{document}